\documentclass[conference]{IEEEtran}
\IEEEoverridecommandlockouts

\usepackage{cite}
\usepackage{amsmath,amssymb,amsfonts}
\usepackage{algorithmic}
\usepackage{graphicx}
\usepackage{textcomp}
\usepackage{xcolor}
\usepackage{orcidlink}
\usepackage{enumitem}
\usepackage{booktabs}
\usepackage{array}
\newcolumntype{Y}{>{\centering\arraybackslash}m{1.6em}}

\usepackage{subcaption}

\usepackage{balance}
\usepackage{siunitx}
\usepackage{expl3,xparse} 
\usepackage[most]{tcolorbox}
\tcbuselibrary{listings} 
\usepackage{cleveref}
\usepackage{listings} 
\usepackage{todonotes}
\usepackage{xspace}

\usepackage{float}

\usepackage{caption}

\makeatletter
\newcommand{\TODO}[1]{%
 \bgroup
 \def\@tempa{#1}%
 \expandafter\textcolor\expandafter{red}{\@tempa}%
 \GenericWarning{}{LaTeX Warning: TODO: \@tempa}%
 \egroup
}
\makeatother

\makeatletter
\newcommand{\NOTE}[1]{%
 \bgroup
 \def\@tempa{#1}%
 \expandafter\textcolor\expandafter{blue}{\@tempa}%
 \GenericWarning{}{LaTeX Warning: NOTE: \@tempa}%
 \egroup
}
\makeatother

\usepackage{soul}

\usepackage{tikz}
\usetikzlibrary{arrows.meta,patterns,calc}
\usepackage{tabularx}
\tcbset{
  pefttile/.style={
    enhanced,
    colback=white, colframe=black,
    rounded corners,
    boxsep=1.5pt, left=3pt, right=3pt, top=3pt, bottom=3pt,
    title style={font=\scriptsize\bfseries},
  }
}

\definecolor{lightgreen}{rgb}{0.85, 1.0, 0.85}
\definecolor{lightred}{rgb}{1.0, 0.85, 0.85}

\newcounter{finding}%[section] % having [section] will make findings numberings local to each section and restart in the upcoming sections
\crefname{finding}{Finding}{Findings} % for \cref
\Crefname{finding}{Finding}{Findings} % for \Cref

\newcommand{\highlightlabelgreen}[2]{%
  \begingroup%
  \sethlcolor{lightgreen}%
  \phantomsection\label{#1}\hl{#2}%
  \endgroup%
}

\newcommand{\highlightlabelred}[2]{%
  \begingroup%
  \sethlcolor{lightred}%
  \phantomsection\label{#1}\hl{#2}%
  \endgroup%
}

\definecolor{lightblue}{RGB}{173, 216, 230}
\newcommand{\highlightlabelblue}[2]{%
  \begingroup%
  \sethlcolor{lightblue}%
  \phantomsection\label{#1}\hl{#2}%
  \endgroup%
}

\tcbset{
  flamesfinding/.style={
  breakable,
    colback=blue!3!white,
    colframe=blue!50!black,
    coltitle=white,
    fonttitle=\bfseries,
    colbacktitle=blue!70!black,
    enhanced,
    sharp corners=all,
    attach boxed title to top center={yshift=-1mm},
    boxed title style={
        size=small,
        colframe=blue!50!black,
        colback=blue!70!black,
        sharp corners=all,
    },
    drop shadow southeast,
    boxrule=0.2mm,
    arc=0.5mm,
    top=0.7mm,
    bottom=0.7mm,
    left=0.7mm,
    right=0.7mm,
    boxsep=0.5mm,
  }
}

\usepackage{listings, xcolor}

\definecolor{verylightgray}{rgb}{.97,.97,.97}

\lstdefinelanguage{Solidity}{
	keywords=[1]{anonymous, assembly, assert, balance, break, call, callcode, case, catch, class, constant, continue, constructor, contract, debugger, default, delegatecall, delete, do, else, emit, event, experimental, export, external, finally, for, function, gas, if, implements, import, in, indexed, instanceof, interface, internal, is, length, library, log0, log1, log2, log3, log4, memory, modifier, new, payable, pragma, private, protected, public, pure, push, require, return, returns, revert, selfdestruct, send, solidity, storage, struct, suicide, super, switch, then, this, throw, transfer, try, typeof, using, value, view, while, with, addmod, ecrecover, keccak256, mulmod, ripemd160, sha256, sha3}, % generic keywords including crypto operations
	keywordstyle=[1]\color{blue}\bfseries,
	keywords=[2]{address, bool, byte, bytes, bytes1, bytes2, bytes3, bytes4, bytes5, bytes6, bytes7, bytes8, bytes9, bytes10, bytes11, bytes12, bytes13, bytes14, bytes15, bytes16, bytes17, bytes18, bytes19, bytes20, bytes21, bytes22, bytes23, bytes24, bytes25, bytes26, bytes27, bytes28, bytes29, bytes30, bytes31, bytes32, enum, int, int8, int16, int24, int32, int40, int48, int56, int64, int72, int80, int88, int96, int104, int112, int120, int128, int136, int144, int152, int160, int168, int176, int184, int192, int200, int208, int216, int224, int232, int240, int248, int256, mapping, string, uint, uint8, uint16, uint24, uint32, uint40, uint48, uint56, uint64, uint72, uint80, uint88, uint96, uint104, uint112, uint120, uint128, uint136, uint144, uint152, uint160, uint168, uint176, uint184, uint192, uint200, uint208, uint216, uint224, uint232, uint240, uint248, uint256, var, void, ether, finney, szabo, wei, days, hours, minutes, seconds, weeks, years},	% types; money and time units
	keywordstyle=[2]\color{teal}\bfseries,
	keywords=[3]{block, blockhash, coinbase, difficulty, gaslimit, number, timestamp, msg, data, gas, sender, sig, value, now, tx, gasprice, origin},	% environment variables
	keywordstyle=[3]\color{violet}\bfseries,
	keywords=[4]{true},
	keywordstyle=[4]\color{green!50!black},
	keywords=[5]{false},
	keywordstyle=[5]\color{red!80!black},
	identifierstyle=\color{black},
	sensitive=false,
	comment=[l]{//},
	morecomment=[s]{/*}{*/},
	commentstyle=\color{gray}\ttfamily,
	stringstyle=\color{red}\ttfamily,
	morestring=[b]',
	morestring=[b]"
}

\definecolor{precolor}{HTML}{4169E1}    % RoyalBlue
\definecolor{vulncolor}{HTML}{FF4500}   % OrangeRed
\definecolor{postcolor}{HTML}{228B22}   % ForestGreen

\colorlet{precolorlight}{precolor!25}
\colorlet{vulncolorlight}{vulncolor!25}
\colorlet{postcolorlight}{postcolor!25}

\newcommand{\conditionDotBox}[3]{%
  \begingroup
  \renewcommand{\arraystretch}{0}%
  \setlength{\tabcolsep}{0pt}%
  \begin{tabular}{@{}c@{}}%
    \colorbox{precolorlight}{\makebox[0.4em][c]{\ifnum#1=1\textcolor{black}{$\bullet$}\else\phantom{$\bullet$}\fi}} \\%
    \colorbox{postcolorlight}{\makebox[0.4em][c]{\ifnum#3=1\textcolor{black}{$\bullet$}\else\phantom{$\bullet$}\fi}}%
  \end{tabular}%
  \endgroup
}

\newcommand{\legendDotItem}[2]{%
    \colorbox{#1}{\makebox[0.4em][c]{\textcolor{black}{$\bullet$}}}\hspace{0.5em}#2%
}
\usepackage{array,booktabs}
\newlength{\TripColW}
\newlength{\ModelSep}
\newlength{\ModelBlockW}
\newcolumntype{B}{>{\centering\arraybackslash}m{\ModelBlockW}}

\newcommand{\trip}[3]{%
  \begingroup\setlength{\tabcolsep}{2pt}\renewcommand{\arraystretch}{1}%
  \begin{tabular}{@{}m{\TripColW}m{\TripColW}m{\TripColW}@{}}
    \centering #1 & \centering #2 & \centering #3
  \end{tabular}%
  \endgroup
}

\newcommand{\hdr}[1]{%
  \begingroup\setlength{\tabcolsep}{0pt}\renewcommand{\arraystretch}{1}%
  \begin{tabular}{@{}c@{}}\textbf{#1}\\[-1pt]\rule{0.98\ModelBlockW}{0.4pt}\end{tabular}%
  \endgroup
}

\newcommand{\flames}{\textsc{Flames}\xspace}
\newcommand{\disl}{\textsc{Disl}\xspace}
\newcommand{\sindi}{\textsc{Sindi}\xspace}
\newcommand{\cl}{\textsc{CodeLlama}\xspace}

\newcommand{\hardinv}{\textsc{Disl-HardInv}\xspace}

\DeclareRobustCommand{\code}[1]{\lstinline[basicstyle=\ttfamily\small]{#1}}

\usepackage{inconsolata}
\DeclareRobustCommand{\code}[1]{\lstinline[basicstyle=\ttfamily\small]{#1}}

\def\BibTeX{{\rm B\kern-.05em{\sc i\kern-.025em b}\kern-.08em
    T\kern-.1667em\lower.7ex\hbox{E}\kern-.125emX}}
\begin{document}

\title{\flames: Fine-tuning LLMs to Synthesize Invariants for Smart Contract Security }

\author{
\IEEEauthorblockN{
Mojtaba Eshghie\orcidlink{0000-0002-0069-0588}\textsuperscript{1},
Gabriele Morello\textsuperscript{2},
Matteo Lauretano\textsuperscript{2},
Alexandre Bartel\textsuperscript{1},
Martin Monperrus\textsuperscript{2}
}
\\
\IEEEauthorblockA{\textsuperscript{1}\textit{Umeå University}, Umeå, Sweden}
\\
\IEEEauthorblockA{\textsuperscript{2}\textit{KTH Royal Institute of Technology}, Stockholm, Sweden}
\thanks{Emails: \{mojtabae, alexandre.bartel\}@cs.umu.se; gabriele.morello.123@gmail.com, matteola@kth.se, monperrus@kth.se.}
}

% \author{
% \IEEEauthorblockN{
% Mojtaba Eshghie\orcidlink{0000-0002-0069-0588}\textsuperscript{1},
% Gabriele Morello\textsuperscript{2},
% Matteo Lauretano\textsuperscript{2},
% Alexandre Bartel\textsuperscript{1},
% Martin Monperrus\textsuperscript{2}
% }
% \IEEEauthorblockA{\textsuperscript{1}\textit{Department of Computing Science, Umeå University}, Umeå, Sweden \\
% Emails: \{mojtabae, alexandre.bartel\}@cs.umu.se}
% \IEEEauthorblockA{\textsuperscript{2}\textit{KTH Royal Institute of Technology}, Stockholm, Sweden \\
% Emails: gabriele.morello.123@gmail.com, matteola@kth.se, monperrus@kth.se}
% }

% \author{\IEEEauthorblockN{Anonymous Authors}}

\maketitle

\begin{abstract}
Smart contract vulnerabilities cost billions of dollars annually, yet existing automated analysis tools fail to generate deployable defenses. We present FLAMES, a novel automated approach that synthesizes executable runtime guards as Solidity require statements to harden smart contracts against exploits. Unlike prior work that relies on vulnerability labels, symbolic analysis, or natural language specifications, FLAMES employs domain-adapted large language models trained through fill-in-the-middle supervised fine-tuning on real-world invariants extracted from 514,506 verified contracts.
Our extensive evaluation across three dimensions demonstrates FLAMES's effectiveness: (1) Compilation: FLAMES achieves 96.7\% compilability for synthesized invariant (2) Semantic Quality: on a curated test set of 5,000 challenging invariants, FLAMES produces exact or semantically equivalent matches to ground truth in 44.5\% of cases; (3) Exploit Mitigation: FLAMES prevents 22 out of 108 real exploits (20.4\%) while preserving contract functionality, and (4) FLAMES successfully blocks the real-world APEMAGA incident by synthesizing a pre-condition that mitigates the attack.
FLAMES establishes that domain-adapted LLMs can automatically generate production-ready security defenses for smart contracts without requiring vulnerability detection, formal specifications, or human intervention. We release our code, model weights, datasets, and evaluation infrastructure to enable reproducible research in this critical domain.
\end{abstract}


\section{Introduction}

Smart contracts underpin blockchain applications for managing digital assets. Despite the progress in automated smart contract analysis, billions of dollars are lost every year due to exploits abusing smart contract vulnerabilities~\cite{SoK,zhou_sok_2023}. Most smart contract  vulnerability detectors produce unreliable verdicts and  do not generate deployable defenses~\cite{HowEffectiveAreSCTools,SCDeFiSecTools,sb_heist_paper}. As a result, there is strong demand for practical techniques to generate concrete security mitigations.

A promising solution is \emph{invariant-based} defense: \emph{pre-/post-conditions} %and \emph{sensitive line} checks (e.g., before external contract calls) 
that revert dangerous executions at runtime~\cite{DesignByContract}. This is implemented in the widely used \code{require} statement in the Solidity language for smart contracts ~\cite{EthereumYellowPaper2022}. Recent work further confirms that attacks are preventable by invariants~\cite{DemystifyingInvEffectiveness}. 
However, automatically obtaining \emph{useful} invariants for real contracts is a hard open research problem. Dynamic invariant mining from on-chain transaction traces 
can surface likely properties~\cite{AutoInvGenSC}, but are noisy and miss important security aspects~\cite{DemystifyingInvEffectiveness}. 
%On another note, analysis tools detect vulnerabilities but do not provide ready-to-inject invariants as pre-/post-conditions~\cite{echidna,HowEffectiveAreSCTools}. 

% our technical contribution
Our core insight is to use Large Language Models (LLM) to synthesize invariants. Yet, off-the-shelf code LLMs are not good at handling Solidity smart contracts without domain adaptation~\cite{SolBench,SolEval,TrailOfBits}.
We propose \flames\footnote{\url{https://github.com/ASSERT-KTH/FLAMES}}, a novel training and inference pipeline to synthesize high-quality invariants as Solidity pre-/post-conditions.
\flames\ (i) analyzes the contract to extract a contract context to be given as input to the LLM; (ii) \flames uses \emph{fill-in-the-middle} supervised fine-tuning (SFT) over real-world \code{require(<invariant>)} statements to train the model to reconstruct missing invariants; and (iii) at inference, \flames\ synthesizes \code{require} invariant statements at function entry (pre-conditions) and exit (post-conditions) locations. %(e.g., before external calls) 
Crucially, \flames\ does \emph{not} require vulnerability labels, vulnerability detectors, or natural-language description, it does fully automated invariant synthesis for contract hardening. 

% novelty
%\todo{one paragraph about the closest related work and our core novelty}
Prior work have mined invariants from transaction traces~\cite{AutoInvGenSC,DemystifyingInvEffectiveness}, used symbolic analysis to infer invariants~\cite{VERISMART}, or prompted LLMs with vulnerability labels to propose invariants~\cite{SMARTINV,PropertyGPT,trustllm}. In contrast, \flames\ generates \emph{concrete}  executable invariants as Solidity \code{require} statements; with no need for specifications or vulnerability labels or hints.
%Beyond demonstrating improved compilability and semantic alignment, we provide end-to-end evidence that these synthesized guards {block real exploits}, and we analyze how placement (pre vs.\ post) and inference strategy (single vs.\ multi-turn) impact effectiveness.

\begin{figure}[!t]
    \centering
    \includegraphics[width=1\columnwidth]{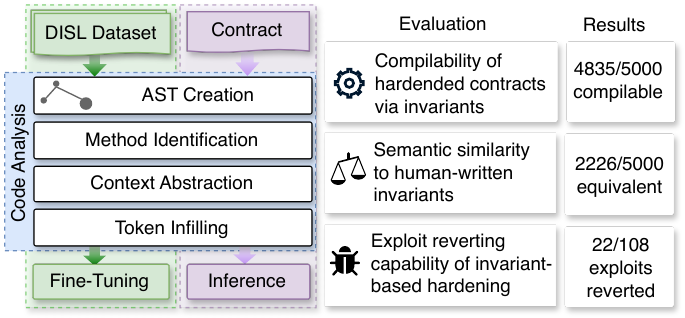}
    \caption{\flames\ architecture, evaluation, and results overview.}
    \label{fig:flames_overview}
\end{figure}

%  evaluation
%Our extensive experimental evaluation is guided by three research questions:
Three research questions guide our extensive experimental evaluations:

\begin{itemize}[leftmargin=8pt]
    \item \textbf{RQ1 (Compilation).} What percentage of invariants synthesized by \flames\ compile when injected into real contracts? \textbf{Answer:} up to {96.7\%} synthesized invariants compile (vs.\ {62.6\%} for the non–domain-adapted baseline). 
    \item \textbf{RQ2 (Semantic Quality).} How close are the synthesized invariants to human-written ones? \textbf{Answer:} on a 5k hard-invariant set, \flames\ attains {44.5\%} exact-or-equivalent matches to ground truth (vs.\ {23.5\%} baseline).%, with exact textual matches rising from {959} to {1840}.
    \item \textbf{RQ3 (Exploit Mitigation).} To what extent do injected invariants prevent exploits while preserving functionality? \textbf{Answer:} the best configuration prevents {22/108} exploits without negatively affecting contract's functionality (vs {16/108} baseline).

    % \item \textbf{RQ4 (Real-World Incident Hardening Case).} How effective \flames\ is in defending against recorded real-world incidents using invariants? \textbf{Answer:} \flames-100k generates an invariant that reverts the attack when synthesizing a pre-condition for a function that is called as an strategic step of the \code{APEMAGA} incident.  

    \item \textbf{RQ4 (Real Incident Hardening).} How effective is \flames at synthesizing a deployable invariant in blocking a successful attack? \textbf{Answer:} In a reproduction of the real-world \code{APEMAGA} incident, \flames-100k synthesizes a pre-condition semantically equivalent to the ground-truth patch, reverting the attack transaction.

\end{itemize}

To sum up, our contributions are:

\begin{itemize}[leftmargin=10pt]
    \item We introduce {\flames}, a novel pipeline for invariant-based smart contract hardening, based on LLM fine-tuning. \flames\ generates valuable pre-/post-conditions using only source code as input.
    \item We curate {\disl}, a dataset of \num{514506} unique, verified Solidity files and a 5k \hardinv test set for rigorous evaluation.
    \item We deliver a comprehensive evaluation demonstrating the high {compilability}, {semantic quality} , and exploit mitigation effectiveness of \flames invariants.
    \item We release our code, model weights\footnote{\url{https://huggingface.co/ASSERT-KTH/FLAMES-100k-2406}}, and evaluation harnesses for reproducibility and future research in this area.
\end{itemize}

% Our experiments provide the first large-scale evidence that specialized fine-tuning enables practical, invariant-based hardening of smart contracts.

%The rest of the paper is structured as with 
\Cref{sec:flames} describes the design of \flames, \Cref{sec:disl} expands the details of the collection, deduplication, and contents of our \disl\ dataset, \Cref{sec:experimental_setup} describes the experimental protocol to answer RQ1--RQ4, \Cref{sec:results} provide our results to the RQs, \Cref{sec:related-works} provides an account of the most relevant scientific papers, \Cref{sec:threats_to_validity} discusses the threats to validity, and \Cref{sec:conclusion} concludes the paper.

\begin{figure}[!t]
    \centering
    \includegraphics[width=0.6\columnwidth]{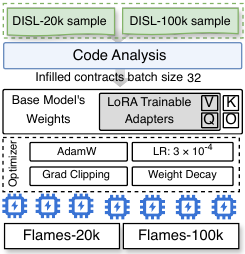}
    \caption{The supervised fine-tuning pipeline yields to \flames models, specialized for Solidity.}
    \label{fig:ft_pipeline}
\end{figure}

\section{\flames Architecture}\label{sec:flames}

We propose
\flames, a novel smart contract hardening pipeline via specialized LLMs.
\flames follows a multi-stage process: a code analysis stage (\Cref{sec:code_analysis}), a one-time supervised fine-tuning stage (\Cref{sec:ft}), and an inference stage (\Cref{sec:inference}) to synthesize missing invariants in smart contracts.

\subsection{Overview}
\label{sec:overview}
The core idea of \flames\ is to inject invariants to harden smart contracts.
Invariants are boolean conditions that must hold at specific program points, such as pre-conditions at function  entry and post-conditions at function exit. In Solidity, they are written with \code{require(<invariant>)} statements. When a require statetement is evaluated as false, transaction execution reverts, preventing any state changes with guarantees. %Local, cheap, and composable, such guards turn security assumptions into executable checks.

%\textit{Example (BatchOverflow, 2018).}
In April 2018, the BeautyChain contract suffered a attack abusing an overflow vulnerability in its code~\cite{BECIncident}. As \Cref{fig:BEC} shows, BeautyChain implemented a \code{batchTransfer} that computed \code{amount = _value * _rec.length} {before} any overflow check. The attacker exploited this publicly callable vulnerable function by sending a transaction with a very large \code{_value} resulting in an overflow and wrap around letting attackers mint gigantic balances. The one-line pre-condition invariant in \Cref{line:overflow-guard} would have blocked the attack transaction. Invariants at the right point can stop exploits mid-flight.

\begin{figure}
    \centering
\begin{lstlisting}[language=Solidity,escapeinside={(*@}{@*)}]
function batchTransfer(address[] _rec, uint256 _value) public whenNotPaused returns (bool) {
  (*@\highlightlabelgreen{line:overflow-guard}{require(\_rec.length == 0||\_value <= type(uint256).max/\_rec.length);}@*)
  uint cnt = _rec.length;
  uint256 amount = uint256(cnt) * _value;
  require(cnt > 0 && cnt <= 20);
  require(_value > 0 && balances[msg.sender] >= amount);
  balances[msg.sender] = balances[msg.sender].sub(amount);
  for (uint i = 0; i < cnt; i++) {
    balances[_rec[i]] = balances[_rec[i]].add(_value);
    Transfer(msg.sender, _rec[i], _value);
  }
  return true;
}
\end{lstlisting}
    \caption{Preventing the BeautyChain attack with a pre-condition invariant.}
    \label{fig:BEC}
\end{figure}

\Cref{fig:flames_overview} shows our proposed solution, called  \flames. It follows a three-stage pipeline to harden smart contracts without any prior vulnerability information: {(i) code analysis (\S\ref{sec:code_analysis})}, {(ii) supervised fine-tuning (\S\ref{sec:ft})}, and {(iii) invariant synthesis (\S\ref{sec:inference})}. Given a contract, the code-analysis stage parses the source into an AST, determines the invariant synthesis location, builds a compact version of the contract and then, the extracted data is used to fine-tune the LLM or to synthesize the invariants.

\textbf{Fine-tuning.} During the fine-tuning stage, we mine real-world invariants of the form of \code{require(...)} statements and conduct supervised fine-tuning to obtain a model specialized for generating smart-contract invariants.

\begin{figure}[!t]
    \centering
    \includegraphics[width=1\columnwidth]{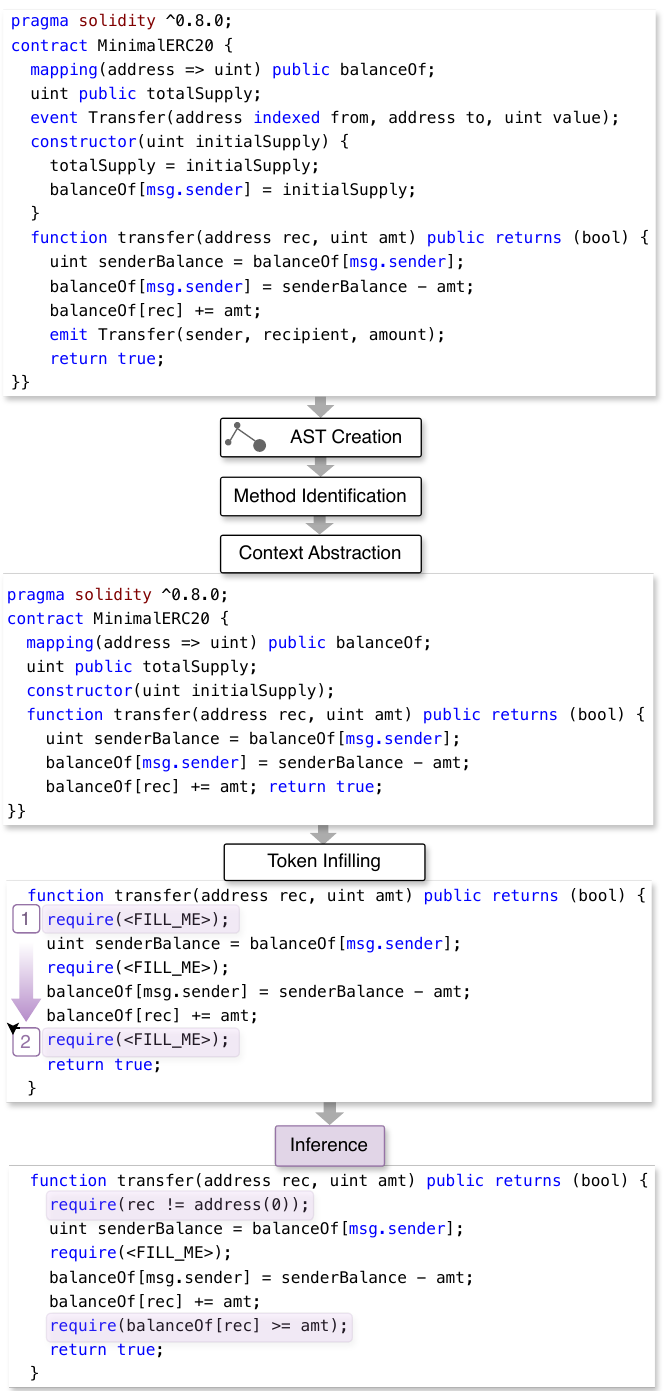}
    \caption{\flames\ synthesis stage illustrated with an example.}
    \label{fig:inference-pipeline}
\end{figure}

\textbf{Inference.} At inference time (\Cref{fig:inference-pipeline}), a user specifies one or more synthesis locations and asks the model to generates the missing invariant(s).   
The synthesized invariants are then injected into the full source in order to permanently harden the contract.

\subsection{Code Analysis Stage}\label{sec:code_analysis}

In \flames, the same code analysis is applied at training time, to prepare the data, and at inference time, to synthesize new invariants.
\flames\ first parses the smart contract into an Abstract Syntax Tree (AST). The AST representation helps manipulate the code programmatically~\cite{SoliDiffy}.  
% abstraction
%\todo[inline]{explain in one full paragraph te problem, give number: XX\% of contracts do not fit in model X or Z}

\begin{figure}[!t]
    \centering
    \includegraphics[width=0.75\columnwidth]{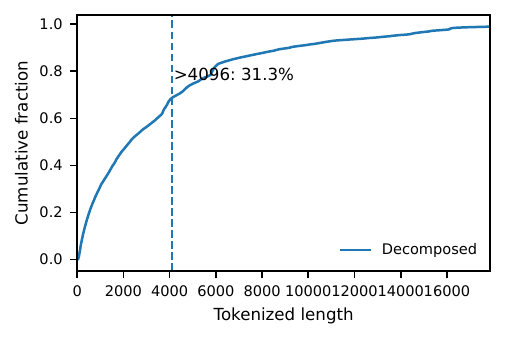}
    \caption{Cumulative distribution of the token length of \disl contracts.}
    \label{fig:cdf_token}
\end{figure}

A key challenge with using LLMs both during the fine-tuning and inference stage is their limited context window. Indeed, many real-world smart-contracts do not fit into this context, which is a blocker for both fine-tuning and inference. As the token length distribution of the sources in \disl in \Cref{fig:cdf_token} shows, more than \num{31}\% of the dataset overflows \num{4096} which is the maximum token length of many base LLM models. This is a major problem for both training and inference This token length typically represents \(\approx\)\num{300} lines of code for Solidity smart contracts.  
% \todo{update: per our dicussion \Cref{fig:hexbin_tokens} shows the relationship between the tokenized length of contracts in \disl\ %(when tokenized with \cl\ tokenizer) 
% and the number of source code lines. It demonstrates a near-linear relationship between non-empty LoC (y-axis) and tokenized length (y-axis).  The dashed vertical line marks our maximum \num{4096}-token context budget~\cite{Repairllama}.  Its intersection with the dense ridge occurs \(\approx\)\num{300} LoC, meaning that contracts start to overflow the context window beyond %$\sim$
% \num{300} lines. Aggregated over the whole \disl, \num{31.3}\% of files exceed \num{4096} tokens.}
% (i.e., lie to the right of the dashed line). 
%This means roughly one-third of real, unique verified Solidity files would not fit into a standard 4k-token window. 
To handle contracts more than \num{4096} tokens, we employ a \emph{context abstraction} technique adapted from prior work~\cite{sequencer}.
It consists of only keeping the most important parts of the code relevant to the task at hand. \flames abstracts the smart contract as follows: 1) it retains the full body of the function containing the current invariant (to be learned or to be synthesized at training and inference time respectively);
2) it retains all function definitions that call the target function,
3) it keeps modifier definitions applied on the mentioned functions  
4) it only keeps signature of the other functions.
5) it keeps state variable declarations, 
and
6) it removes event definition.
All this is done precisely based on AST manipulation.

\emph{Analysis at training time.} \flames\ conducts contract analysis to prepare the code for the fine-tuning pipeline. It first extracts the \code{require} statements to be fine-tuned on. 
Given the \code{require} statements, it determines their enclosing function and prepares the abstract contract context based on the identified function. The rest of the procedure is described in \autoref{sec:ft}.

\emph{Analysis at inference time.} \Cref{fig:inference-pipeline} shows an instance of this code analysis abstraction during the inference stage. \flames only keeps the state variable declarations \code{balancedOf} and \code{totalSupply}, preserves the signature of the \code{constructor}, and the full definition of the \code{transfer} function which is target of invariant synthesis.

% \begin{figure}[!t]
%     \centering
%     \includegraphics[width=0.75\columnwidth]{figures/hexbin_tokens_vs_loc_guides_cropped.pdf}
%     \caption{Token length vs. LoC in \disl (99th percentile crop).}
%     \label{fig:hexbin_tokens}
% \end{figure}

\subsection{Supervised Fine-Tuning Stage}\label{sec:ft}

\flames specializes a base LLM for invariant generation via supervised fine-tuning. Each training sample is built from a real \code{require(<invariant>)} taken from the training dataset (see \S\ref{sec:datasets}).
The training sample is generated as follows
(i) we use the abstract context of the contract (\S\ref{sec:code_analysis});
(ii) we replace the invariant with the placeholder token \code{<FILL\_ME>} and keep the surrounding code intact.
This is the \emph{fill-in-the-middle} (FIM) training objective~\cite{FIMOpenAI}. 
Each invariant inside a \code{require} statement in the training dataset is meant to be recreated at the FIM location, based and the surrounding contract.

% The FIM token are at function entry/exit or at the target line \todo{never explained what is a "vulnerable line", never explained what is the ground truth, esp at training time}. 

% \todo{explain / define the concept of "Defensive predicate synthesis"}Defensive predicate synthesis via SFT benefits from domain adaptation (from real-world decentralized applications written in Solidit) while preserving the base model's general code competence. 

%\paragraph*{Memory efficiency via 4-bit loading (QLoRA).}

\subsection{Invariant Synthesis Stage}\label{sec:inference}

During the inference stage (\Cref{fig:inference-pipeline}), a user provides a full smart contract and specifies a target function for smart-contract synthesis. \flames\ extracts this function, then prepares the abstracted version of the contract keeping the identified function with its full implementation.
It then infills the function by \code{require(<FILL_ME>)} statement at the desired invariant locations, one location to be inferred at a time. The \flames model then generates the boolean predicate to fill the mask. Finally, \flames\ replaces the placeholder with the synthesized invariant and reconstructs the full, hardened smart contract code. Then, it compiles and returns the hardened contract to the user.

% \flames\ supports synthesis in form of \code{require} statements

\subsubsection{Invariant Synthesis Locations}\label{sec:injection_strategies}
\flames considers two places for synthesis, where invariants can revert harmful transactions: pre-condition (function entry point) and post-condition (function exit point), %, and sensitive line(s) (e.g., before/after external \code{call}s/\code{delegatecall}s to other contracts), 
\flames\ supports synthesizing invariants at the mentioned locations.
%seven different combinations of injection locations including at each of them in isolation, selection of two of the three, and all three locations in the target function. 

\subsubsection{Inference Strategies}\label{sec:inference_strategies}

\textit{Single-turn}: When injecting  multiple invariants in the contract, on can prompt the model each time with the original version of the contract.

\textit{Multi-turn}: When generating multiple invariants (e.g., pre- and post-condition in the same contract or function), synthesized invariants are kept in the context for subsequent generations. In other words, if the synthesis task consists of generating multiple pre-/post-conditation, the model will be prompted multiple times, and at each turn, it is given the contract containing the newly synthesized invariants from previous turns~\cite{SelfRefine,CodeGen}.

\subsection{Implementation}
% We load the frozen base model \textsc{CodeLlama2-7B} in 4-bit quantization and train LoRA parameters in 16-bit mixed-precision~\cite{lora,QLORA}. This preserves the representational capacity of the base model while cutting activations enough to fine-tune on commodity multi-GPU machines. This combination (i) lowers the training memory footprint so we can use larger training corpora; (ii) constrains the update to a small subspace of the whole model. The details of fine-tuning with this training regime are provided in \Cref{sec:sft_appendix}.
% We use AdamW, gradient accumulation to reach the effective batch size in \S\ref{sec:experimental_setup}, and standard stability techniques (gradient clipping). 

% \todo{we use PEFT~\cite{PEFTRootPaper} but we don't mention anything for it. Write it cohesively with the rest of the things in this subsection}

We implement our fine-tuning stage using the  Parameter-Efficient Fine-Tuning (PEFT) library of Huggingface~\cite{PEFTRootPaper}. %, which provides a  framework for parameter-efficient adaptation methods. 
We use \textsc{CodeLlama2-7B} as base model, in 4-bit quantization.
We employ the QLoRA technique and LoRA adapters for fine-tuning. 
This parameter-efficient fine tuning (i) reduces the training memory footprint enabling us to use larger training corpora on commodity multi-GPU machines; (ii) constrains parameter updates to a low-rank subspace while preserving the base model's capabilities.
We use AdamW optimizer with a learning rate of $3 \times 10^{-4}$  and a batch size of 32 (with two batches per device), and gradient clipping for stability. %We used four A100 NVIDIA GPUs for the fine-tuning stage. 
 \Cref{tab:finetuning_flames} details the system setup for the fine-tuning stage.
The training runs for one epoch on the selected subset of the \disl\ dataset (see \S\ref{sec:experimental_setup}).
We fine-tune two versions of \flames, one with \num{20000} samples (\flames-20k), another with \num{100000} samples (\flames-100k).

\begin{table}
\caption{Training parameters of supervised fine-tuning for \flames}
\label{tab:finetuning_flames}
\centering
\scriptsize
\begin{tabular}{p{1.55cm}p{6.5cm}}
\toprule
\multicolumn{2}{c}{\textbf{Hardware and Runtime}}\\
\midrule
\textbf{Train GPUs} & $4\times$ Nvidia A100 (80\,GB each) \\
\textbf{Precision} & 4-bit quantized load (nf4) with mixed precision (fp16) for training \\
\textbf{Optimizer} & AdamW \\
\textbf{Batch size} & 32 (per-device batch: 2; gradient accumulation: 16) \\
\textbf{Epochs} & 1 (for both \flames-20k and \flames-100k) \\
\textbf{Learning rate} & $3\times 10^{-4}$ \\
\addlinespace
\multicolumn{2}{c}{\textbf{Model and PEFT Configuration}}\\
\midrule
\textbf{Base model} & CodeLlama2-7B \\
\textbf{LoRA} & rank $r{=}8$, $\alpha{=}16$; injected in attention Q/V projections \\
\textbf{Context length} & \num{4096} tokens \\
\textbf{Infilling mode} & Fill-in-the-Middle (FIM) with \texttt{<FILL\_ME>} \\
\bottomrule
\end{tabular}
\end{table}

\section{DISL: Fine-Tuning Dataset}
\label{sec:disl}

A major challenge in applying machine learning to smart contract security is the lack of unique large datasets of real-world contracts. Recent studies indicate that synthetic, small, or non-diverse datasets are not satisfactory~\cite{HowEffectiveAreSCTools, ISSTA2021EmpiricalEvaluation}. 

To address this gap, we create \disl, the largest dataset of \emph{unique} Solidity smart contracts that serves as the foundation for training and evaluating \flames. \disl\ is designed to be diverse, recent, and suitable for AI tasks. It focuses exclusively on real-world contract with verified source code.

\subsection{Data Collection}

To ensure clarity, we define the following terms:
\begin{itemize}[leftmargin=8pt]
    \item {Deployed contract}: A smart contract address on the Ethereum mainnet associated with executable bytecode.
    \item {Raw contract}: The complete, concatenated source code retrieved from Etherscan for a single deployed contract, which may include multiple files.
    \item {Solidity file}: A single ``.sol'' source file, which may itself contain multiple \code{contract} or \code{library} definitions.
\end{itemize}

Our collection process was executed in two phases:

\paragraph{Initial Seeding and Expansion} We began with the Storhaug et al.'s dataset \cite{andstorHuggingface,andstorPaper}, which contains \num{2217692} raw contract records up to April 1, 2022. To capture more recent contracts, we queried the Google BigQuery Ethereum database for all contracts deployed between April 1, 2022, and January 15, 2024, that had at least one on-chain transaction~\cite{EthereumBigQuery}. This query yields \num{2709030} contract addresses.

\paragraph{Source Code Retrieval} Using the Etherscan API, we fetched the verified source code for these addresses. A contract is ``verified'' if its uploaded source compiles to the exact bytecode on the blockchain, ensuring authenticity. After filtering out unverified contracts, this step yielded \num{1080579} new raw contract records. Merging these with the seed dataset resulted in a total raw collection of \num{3298271} \emph{deployed contracts}.

After this curation, we obtain the final, unique \disl\ dataset. 
\disl\ consists of two collections: i) decomposed set containing \num{12931943} Solidity files which is the result of splitting each raw contract record into its individual Solidity files. This process expanded our collection to a total of \num{12931943} Solidity files. and ii) the decomposed and duduplicated (unique) set consisting of \num{514506} Solidity files: to eliminate redundant code, we used the Jaccard similarity index with a \num{90}\% threshold to compare the sets of tokens in each file~\cite{adverseEffectsOfDuplicateCode}. This way, we identify and remove near-duplicates and preserve unique implementations. This step was critical as it revealed that over \num{96}\% of the decomposed files were duplicates. 

Each entry of \disl\ is enriched with valuable metadata (see repository), which is crucial for tasks requiring compilation and metadata analysis, as we do in \flames. As \Cref{tab:datasets} shows, \disl\ is the largest Solidity smart contract dataset to date based on number of files, recency, uniqueness, and diversity.

Our analysis shows the deduplicated subset of \disl\ contains \num{5276315} \code{require} statements. We use these \code{require} statements for both the supervised fine-tuning and evaluation of \flames\ (RQ1 and RQ2), with seggegration to avoid data leakage.
%Both the raw and decomposed collections are publicly available in Parquet format on Hugging Face: \url{https://huggingface.co/datasets/ASSERT-KTH/DISL}.

\begin{table}[t]
    \centering
    \caption{The other smart contract datasets. \disl is foundational for future learning-based approaches on smart contracts.}
    \renewcommand{\arraystretch}{1.05}
    \begin{tabular}{p{4.5cm}p{1cm}>{\raggedleft\arraybackslash}p{1.8cm}}
        \toprule
        \textbf{Dataset} & \textbf{Year} & \textbf{Size} \\
        \midrule
        Storhaug (\emph{deduplicated})~\cite{andstorHuggingface} & 2022 & \num{186397} \\
        Fiesta\textsuperscript{1} (\emph{deduplicated})~\cite{ZellicHuggingface} & 2023 & \num{149386} \\
        Sanctuary~\cite{smart_contract_sanctuary} & 2022 & \num{144857} \\
        SmartBugs-Wild~\cite{SBWild,ThomasEmpricalReview} & 2020 & \num{47587} \\
        Ren et al. ~\cite{ISSTA2021EmpiricalEvaluation,SCBenchmarkSuitesUnified} & 2021 & \num{46186} \\
        DAppSCAN~\cite{DAppSCAN} & 2023 & \num{39904} \\
        \midrule % Replaced \hline with \midrule for consistency
        \textbf{\disl\ (\emph{decomposed})} &  & \textbf{\num{12931943}} \\
        \textbf{\disl\ (\emph{decomposed \& deduplicated})} &  & \textbf{\num{514506}} \\
        \bottomrule
    \end{tabular}
    \label{tab:datasets}
    \begin{minipage}{8.5cm} % Adjust the width to match your table width
        \scriptsize
        \vspace{0.5em} % Optional: adds a little space before the note
        1. Uses exact match for deduplication
    \end{minipage}
\end{table}

\section{Experimental Evaluation}\label{sec:experimental_setup}

We design a comprehensive set of experiments to evaluate \flames, addressing the following research questions:

\begin{itemize}[leftmargin=8pt]
    \item \textbf{RQ1 Compilation:} To what extent do invariants synthesized by \flames real smart contracts compile when injected back in the code?
    \item \textbf{RQ2 Semantic Quality:} How semantically close are the \flames invariants compared to human-written ones?% (exact match / semantic equivalence / relative semantic strength)?
    \item \textbf{RQ3 Mitigation:} To what extent do \flames invariants prevent exploits while preserving functionality?
    \item \textbf{RQ4 (Real Incident Hardening).} How effective is \flames\ at synthesizing a deployable invariant to block a real-world incident?
\end{itemize}

This section details the datasets (\Cref{sec:datasets}) and the experimental protocols for RQ1--RQ3 (\Cref{sec:rq1_protocol,sec:rq2_protocol,sec:rq3_protocol}).

% \todo{give the research questions here}

\subsection{Datasets Used for Experiments}\label{sec:datasets}

\subsubsection{Fine Tuning Datasets} 
    
We train \flames{}-20k and \flames{}-100k on \num{20000} and \num{100000} samples, respectively, from the \disl\ deduplicated set. Each sample is a full contract file from \disl\ with one of its invariants randomly selected for the supervised fine-tuning stage.

\subsubsection{Benchmark for RQ1 \& RQ2} 

To rigorously code generation quality, we create a challenging test set by filtering \textsc{DISL} and selecting \num{5000} unique, non-trivial invariants.
We do so by removing the ones that contain only variable identifiers, simple comparisons, negations, straightforward function calls, or hardcoded addresses. To prevent data leakage, they are not picked from the fine-tuning corpus. %These \code{require} statements belong to contracts from \num{3589} different unique deployments on Ethereum.  
% https://github.com/ASSERT-KTH/FLAMES/tree/master/Disl-hardinv
We publish \hardinv\ on our repository\footnote{\url{https://github.com/ASSERT-KTH/FLAMES/tree/master/Disl-hardinv}}.

\subsubsection{Benchmark for RQ3} To evaluate real-world effectiveness, we use \num{108} vulnerable contracts from the SB-Heist benchmark~\cite{bobadilla2025automatedfixestrulymitigate}, which are written in Solidity and come with ground-truth exploits. We chose this benchmark since it provides read-to-use proof-of-concept exploits and functionality tests that would capture if hardening has changed the nominal behavior of the contract.

\subsection{Protocol for RQ1}\label{sec:rq1_protocol}
We evaluate the compilability of the synthesized invariants on \hardinv\ dataset (see \S\ref{sec:datasets}). Then, we replace all invariants in \hardinv with \code{<FILL_ME>} token, and infer the missing invariant. We use \cl\ (as baseline), \flames-20k, and \flames-100k to recreate the invariant.  
We then compile the contract, using its original compiler configuration. To prevent data leakage, these contracts are distinct from the fine-tuning dataset presented in \Cref{sec:datasets}.

We report the number of successful compilations for each model. A build \emph{succeeds} if, under the contract's original compiler configuration (pragma/version), \texttt{solc} compiler exits without errors and produces artifacts. A high compilation success post-hardening indicates that the model's synthesized predicates are syntactically valid and are consistent with the surrounding context from a typing, naming and scoping perspective. A high compilation rate is a {necessary} condition for deployable hardening but not a guarantee of semantic correctness or security: those are evaluated separately in RQ2 (semantic relation to human invariants) and RQ3 (exploit prevention).

\subsection{Protocol for RQ2}\label{sec:rq2_protocol}
To assess how well the generated invariants match the human-written ones, we use the \hardinv\ dataset (see \S\ref{sec:datasets}). For each sample, we prompt the models to generate the missing invariant. We then compare the synthesized invariant with the ground-truth using \sindi~\cite{SindiGithub}, a semantic equivalence checker tool for Solidity. \sindi uses a combination of SMT-solving and rewrite rules to determine the semantic relationship of two given invariants. 
Using \sindi, we classify the relationship of the invariant and its ground truth into five categories: exact match, semantically equivalent, synthesized or ground truth stronger, inconclusive.  
For each \(\langle \text{synthesized},\ \text{ground truth}\rangle\) pair, we assign one of five verdicts.

\noindent\emph{Exact Match}. We first compare the invariants in textual form after stripping whitespaces. If they are equal, they are classified as an exact match, otherwise, we move to using \sindi\ for the semantic comparison. 
  
\noindent\emph{Semantically equivalent}. \sindi proves semantic equivalence through mutual logical implication (when syn: synthesized invariant and gt: ground truth then \(\text{syn} \Rightarrow \text{gt}\) and \(\text{gt} \Rightarrow \text{syn}\)):

\noindent\code{__msgSender()} \code{== owner} \(\equiv\) \code{msg.sender == owner}.

\noindent\emph{Synthesized / ground truth stronger}. \sindi proves \(\text{syn} \Rightarrow \text{gt}\) but not the inverse; this means one of the invariants accepts a {smaller} set of states:

\noindent\code{paidCount * mintPrice == msg.value} \(\Rightarrow\) \code{msg.value >= paidCount * mintPrice}.

  % \item \emph{Ground truth stronger}. \sindi proves \(\text{gt} \Rightarrow \text{synth}\) but not the converse; the human-written invariant is stricter.
  % \begin{itemize}[leftmargin=8pt]
  %   \item e.g., \code{amount > 0 \&\& amount <= balances[msg.sender]} \(\Rightarrow\) \code{amount <= balances[msg.sender]}.
  % \end{itemize}

\noindent\emph{Inconclusive}. Neither implication is provable within \sindi's logic or given the solver time budget, or the invariants are over unrelated variables so that no implication holds:

\noindent\code{IERC20(token).balanceOf(msg.sender) >= amount} vs. \code{whitelist[msg.sender]}.

%In practice, ``inconclusive'' arises when predicates constrain disjoint aspects of state (caller identity vs. payment amount); \sindi\ conservatively defaults to this class in such cases.

\subsection{Protocol for RQ3}\label{sec:rq3_protocol}

We use a dataset of vulnerable smart contracts and their exploit  to measure security effectiveness of the synthesized invariants~\cite{sb_heist_paper,sb_heist_repo}. Each one of the \num{108} contracts in this benchmark contains one or more vulnerabilities in one or more lines in the source code.

For each of these vulnerability locations, we synthesize invariants with \emph{three} different injection strategies: pre-condition, post-condition, and both pre and post for the same contract.
%The target line number in this context is the same as the vulnerability location information for each contract. 
For each injection strategy and inference strategy (single- and multi-turn), we synthesize the missing invariants, and inject them at their determined location.

For each candidate hardened version, we then execute: 
\begin{enumerate}[leftmargin=12pt]
    \item The contract's functional tests to check if the invariant introduces any regression.
    \item We run the ground-truth exploit on the hardened contract to determine whether the \flames invariant mitigates the exploit. 
    %see if the exploit transaction is reverted by the injected invariants.  
\end{enumerate}

If the hardened contract compiles, passes all functional tests, and successfully stops the exploit, the hardening task is considered successful.  
%is deemed ``successful''. 
    
%\todo{continue fixing below..}

%\paragraph*{\ds{} configuration (RQ3).}
%\textbf{Baselines. } \todo{against codellama, update DeepSeek}
We benchmark our models against \cl~\cite{codellama}. We evaluate \cl and \flames\ using the exact same harness. This gives an insight about \flames's performance against a non–domain-adapted model on the hardening task. 
The Python notebooks used in our experiments are publicly available\footnote{\url{https://github.com/ASSERT-KTH/FLAMES/blob/master/raw-validation-results/sb-heists/}}.

\subsection{Protocol for RQ4}\label{sec:rq4_protocol}
In this RQ, we check whether \flames is able to prevent a real-world smart contract incident.
We choose an attack  based on two criteria: i) the concrete exploit proof-of-concept is available to reproduce the incident ii) it is in nature preventable using an invariant. After careful review of multiple incidents, we select the \code{APEMAGA} protocol incident that happened in June, 2024 leading to a \(\sim\)9.4~ETH loss. 
The manual review of the public attack transaction\footnote{\href{https://etherscan.io/tx/0x6beb21b53f5b205c088570333ec875b720e333b49657f7026b01ed72b026851e}{APEMAGA attack transaction.}} and the exploit reproduction script~\footnote{\href{https://github.com/SunWeb3Sec/DeFiHackLabs/blob/dc2cf9e53e9ccaf2eaf9806bad7cd914edefb41b/src/test/2024-06/APEMAGA_exp.sol}{DeFiHackLabs Exploit PoC.}} reveals the flow of the attack, and the function(s) where the synthesis of a precise invariant would deter the attack transaction. The patch of the attack is also published and gives us the location of the missing invariant.

We use \flames to synthesize the invariant and put it back in the contract.
We compile the hardened contract.
After hardening, we evaluate the hardening effectiveness by running the exploit PoC on the hardened version with the real state of the blockchain pre-incident. To do this, we fork Ethereum mainnet at block prior the attack transaction, replace the token's runtime at the \code{APEMAGA} address, preserving its storage. Then, we run the exploit script to check the effectiveness of the invariant against the attack.

\begin{figure}[t]
    \centering
    \includegraphics[width=0.85\linewidth]{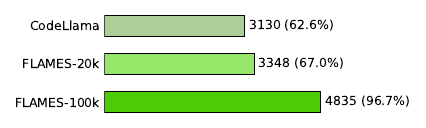}
    \caption{RQ1: Compilation success rate of synthesized invariants for different models.}
    \label{fig:compilation-results}
\end{figure}

\section{Experimental Results}\label{sec:results}

\subsection{RQ1:  Well-formedness of Synthesized Invariants}\label{sec:rq1-res}

We measure whether the \flames synthesized invariants are correct per the syntactic and semantic checks of the compiler.
\Cref{fig:compilation-results} shows the compilability results of 5000 synthesized invariants for the \hardinv\ dataset. The bars show from top to bottom number of contracts compiled without any modification, the number of them compiled after a random \code{require} statement is replaced by its synthesized version, and the two bottom bars pertain to \flames\ models.

As \Cref{fig:compilation-results} shows, our best model \flames-100k, generates \code{require} statements that compile at least 4835/5000 of the times with showing significant improvement over both the baseline and the less fine-tuned variant (\flames-20k). \cl performs poorly with 3130/5000 compiled cases post-hardening, demonstrating the effectiveness of the training loop.

\Cref{fig:codellama-compilation-failure} presents one failed synthesis result where \cl\ continues to generate a function declaration within another function body leading to a parser error during compilation. \cl\ is not able to output a stop token appropriately.
Compared to \cl, \flames-100k does not generate such multi-line generation errors due to incorrect end of generation.

\begin{figure}[H]
\centering

% (a) Failure: CodeLlama drift (use the user's exact snippet)
\begin{subfigure}{0.98\columnwidth}
\centering
\begin{lstlisting}[language=Solidity,escapeinside={(*@}{@*)}]
contract Vault {
  mapping(address => uint256) public balances;  // ... 
  function withdraw(uint256 amount) external {
    (*@\highlightlabelred{line:bad-guard}{require(balances[msg.sender] >= amount);} @*)
    (*@\highlightlabelred{line:bad-helper}{function \_safeSub(uint256 a, uint256 b) internal pure returns}@*)
       (*@\highlightlabelred{line:bad-helper}{(uint256) \{ return a - b; \}}@*)
    balances[msg.sender] -= amount;
    (bool ok, ) = msg.sender.call{value: amount}("");
    require(ok); }  // ...
}
\end{lstlisting}
\caption{Invariant synthesized by \cl\ failing compilation. After infilling the \texttt{require}, \cl\ appends a function \emph{inside} the body of another function, making the contract uncompilable (line breaks added for clarity).}
\label{fig:codellama-compilation-failure}
\end{subfigure}

\medskip

% (b) Success: FLAMES single-predicate, single-line output
\begin{subfigure}{0.98\columnwidth}
\centering
\begin{lstlisting}[language=Solidity,escapeinside={(*@}{@*)}]
contract Vault {
  mapping(address => uint256) public balances; // ...
  function withdraw(uint256 amount) external {
    (*@\highlightlabelgreen{line:good-guard}{require(amount <= balances[msg.sender]);}@*)
    // same as (a)
}
\end{lstlisting}
\caption{\flames\ returns a single well-formed compilable invariant at the mask location.}
\label{fig:flames-compilation-success}
\end{subfigure}

\caption{Synthesized invariants (a) failure by \cl; (b) success by \flames (compiles).}
\label{fig:compilation-behavior-subs}
\end{figure}

% \begin{figure}[H]
% \centering
% \begin{lstlisting}[language=Solidity,escapeinside={(*@}{@*)}]
% contract Vault {
%   mapping(address => uint256) public balances;
%   // ... 
%   function withdraw(uint256 amount) external {
%     (*@\highlightlabelred{line:bad-guard}{require(balances[msg.sender] >= amount);} @*)
%     (*@\highlightlabelred{line:bad-helper}{function \_safeSub(uint256 a, uint256 b) internal pure returns}@*)
%        (*@\highlightlabelred{line:bad-helper}{(uint256) \{ return a - b; \}}@*)
%     balances[msg.sender] -= amount;
%     (bool ok, ) = msg.sender.call{value: amount}("");
%     require(ok); }
%   // ...
% }
% \end{lstlisting}
% \caption{Invariant synthesized by \cl failing compilation. After infilling the \texttt{require}, \cl\ it appends a function {inside} the body of another function, making the contract uncompilable (line breaks are added for clarity)}
% \label{fig:codellama-compilation-failure}
% \end{figure}

\begin{table*}[t]
\centering
\caption{RQ2. Comparison of Ground Truth and Synthesized Invariants, categorized by their logical relationship.}
%\scriptsize
\fontsize{7}{7}\selectfont
\label{tab:synthesized-invariants}
\setlength{\tabcolsep}{3pt} % or 0pt if you want no padding anywhere
\begin{tabular*}{\textwidth}{
  @{}% no left edge padding
  >{\centering\arraybackslash}p{0.5cm}
  @{\hspace{3pt}}% a fixed small gap after the 1cm column (optional)
  >{\raggedright\arraybackslash}p{0.5\textwidth}
  >{\raggedright\arraybackslash}p{0.4\textwidth}
  @{}% no right edge padding
}
\toprule
\textbf{ID} & \textbf{Ground Truth} & \textbf{Synthesized} \\
\midrule
\multicolumn{3}{c}{\textit{\textbf{Equivalent}}} \\
\midrule
1 & \texttt{\_msgSender()==hashesToken.ownerOf(\_hashesTokenId)} & \texttt{hashesToken.ownerOf(\_hashesTokenId)==msg.sender} \\
2 & \texttt{token().transfer(beneficiary,tokensToUnlock)} & \texttt{\_token.transfer(beneficiary,tokensToUnlock)} \\
3 & \texttt{balanceOf[\_to].add(\_value)>=balanceOf[\_to]} & \texttt{balanceOf[\_to]+\_value>=balanceOf[\_to]} \\
4 & \texttt{msg.value>=(cost*(\_mintAmount-1))} & \texttt{msg.value>=((cost*\_mintAmount)-(cost*1))} \\
5 & \texttt{getBlockTimestamp()>=\_payload.valid\_from \&\& getBlockTimestamp()<=\_payload.valid\_to} & \texttt{\_payload.valid\_from<=getBlockTimestamp() \&\& \_payload.valid\_to>=getBlockTimestamp()} \\
\midrule
\multicolumn{3}{c}{\textit{\textbf{Synthesized is stronger}}} \\
\midrule
6 & \texttt{msg.value>=mintPrice*paidCount} & \texttt{paidCount*mintPrice==msg.value}\\
7 & \texttt{msg.sender==gov || msg.sender==ctrl || msg.sender==address(this)} & \texttt{msg.sender==ctrl || msg.sender==gov} \\
8 & \texttt{coinMap[\_c].cContract.balanceOf(msg.sender)>=\_amount} & \texttt{coinMap[\_c].cContract.balanceOf(msg.sender)>=\_amount*1e18} \\
9 & \texttt{offer.offeredTo==address(0x0)||offer.offeredTo==msg.sender} & \texttt{offer.offeredTo==msg.sender} \\
10 & \texttt{block.timestamp<a.allocTime+CANCEL\_PERIOD||a.cancel} & \texttt{a.cancel} \\
\midrule
\multicolumn{3}{c}{\textit{\textbf{Synthesized is weaker}}} \\
\midrule
11 & \texttt{\_userMinted[block.number]<Num() \&\& \_userForFree[tx.origin]<maxFreePerTx} & \texttt{\_userForFree[tx.origin]<maxFreePerTx} \\
12 & \texttt{isEndStage \&\& enterEndTime>0 \&\& now-enterEndTime>END\_STAGE\_DURATION} & \texttt{isEndStage} \\
13 & \texttt{(amount>0) \&\& (GAMintCounter+amount) <= GASupply} & \texttt{amount>0} \\
14 & \texttt{block.timestamp>partyTime \&\& hodlers[msg.sender]>0} & \texttt{now>=partyTime} \\
15 & \texttt{\_redisFeeOnBuy+\_redisFeeOnSell+\_taxFeeOnBuy+\_taxFeeOnSell<=25} & \texttt{\_redisFeeOnBuy+\_redisFeeOnSell + \_taxFeeOnBuy+\_taxFeeOnSell<=50} \\
\midrule
\multicolumn{3}{c}{\textit{\textbf{Inconclusive}}} \\
\midrule
16 & \texttt{(from==owner)||(crowdSalesCompleted>0)} & \texttt{tokens<=allowed[from][msg.sender]} \\
17 & \texttt{\_tx.deadline==0 || \_tx.deadline>block.timestamp} & \texttt{\_tx.to!=address(0)} \\
18 & \texttt{\_raffle.end==0 || block.timestamp>\_raffle.end} & \texttt{\_raffle.isClosed} \\
19 & \texttt{\_maxTransactionAmountBuy>=(totalSupply() / (10**decimals()))/1\_000 \&\& \_maxTransactionAmountSell>=(totalSupply() / (10**decimals()))/1\_000} & \texttt{maxTransactionLimitEnabled} \\
20 & \texttt{payable(teamWallet).send(bal*4/10)\&\&payable(msg.sender).send(bal*2/10) \&\& payable(developmentWallet).send(bal*4/10)} & \texttt{payable(developmentWallet).send(bal)} \\
\bottomrule
\end{tabular*}
\end{table*}

% \todo{never do this: first say what the figure contain,  then, give an example, then, give the finding, this goes far oo fast}

\begin{tcolorbox}[title=RQ1 Answer: Compilability, flamesfinding, label=finding:compilability]
Invariants synthesized by \flames\ compile in \num{96.7} of cases, versus \num{62.6}\% for the baseline. This indicates that domain-specific fine-tuning with \flames\ yields invariants that are of much higher usability.
Fine-tuning enables the model to capture the  syntactic (e.g., scoping) and semantic (e.g., typing) constraints of invariants in the considered smart contract language, Solidity.
%, delivering \(\approx\)5\(\times\) higher compilability than the baseline. 
\end{tcolorbox}

% \todo[inline]{Use the following link for regenerating the plots...}
% \href{https://github.com/ASSERT-KTH/FLAMES/tree/master/raw-validation-results/compilability-results}{https://github.com/ASSERT-KTH/FLAMES/tree/master/raw-validation-results/compilability-results}

\subsection{RQ2: Comparison Against Human-Generated Invariants}\label{sec:rq2-res}

\begin{figure}[!t]
    \centering
    \includegraphics[width=0.8\columnwidth]{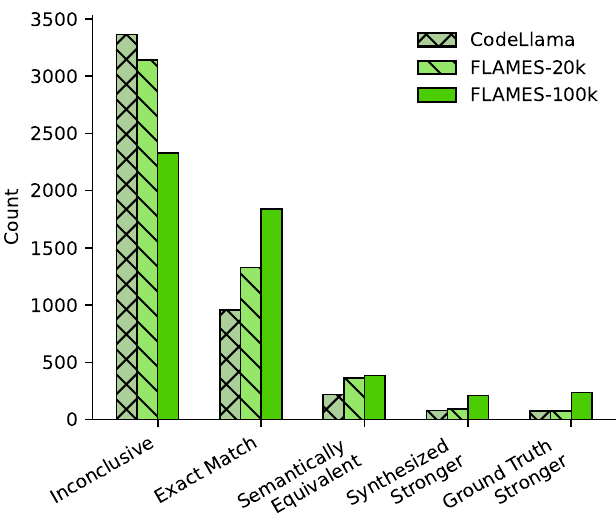}
    \caption{Semantic Similarity of \flames invariants .}
    \label{fig:differencing_results}
\end{figure}

Next, we compare the 5000  synthesized invariants for \hardinv against the original one written by a developer.
\Cref{fig:differencing_results} shows our results for the five categories of semantic relationship between the synthesized invariants and their ground truths for \hardinv.
\flames monotonically improves semantic alignment of synthesized invariants with human-written ones. Moving from the base model to \flames\ trained on \num{20}k and \num{100}k samples monotonically raises \emph{exact} matches (from \num{959} to \num{1328} to \num{1840}). We observe the same improvement via fine-tuning when generating invariants that are textually different but semantically equivalent (from \num{217} to \num{362} to \num{386}).

Fine-tuning also increases the number of cases where the invariant does not semantically or textually equal to the ground truth but still has a semantic relationship (stronger/weaker bars in \Cref{fig:differencing_results}) with its ground truth (\num{157} such cases for \cl\ to \num{168} for \flames-20k to \num{446} cases for \flames-100k). 
Generating invariants that have semantic relationship with their ground truth counts as partial success as stronger synthesized invariants could subsume the allowed behavior of their ground truth and synthesis of multiple weaker invariants (compared to their ground truth) can resolve some security issues.

The number of cases where the invariants are stronger/weaker are balanced for the two categories for all models (e.g., out of \num{446} such cases for \flames-100k, in \num{209} cases the synthesized invariants are stronger and \num{237} vice versa).
\Cref{tab:synthesized-invariants} presents five synthesized invariants (right) from each of the four semantic categories and their ground truth (left).

\textbf{Equivalent Invariants. } 
%\paragraph*{When synthesized is \emph{equivalent} but not textually identical.}
Equivalences arise from algebraic normalization, commutativity, and popular library alias substitutions. In \Cref{tab:synthesized-invariants}, ID~1 swaps sides of equality (\code{\_msgSender() == ownerOf(..)} $\equiv$ \code{ownerOf(..) == msg.sender}) while also replacing the \code{_msgSender()} with the equivalent \code{msg.sender}. ID~3 replaces a SafeMath idiom \code{a.add(b) >= a} with its native form \code{a + b >= a}. ID~4 distributes constants and rewrites \code{>= cost * (\_mintAmount - 1)} as \code{>= (cost * \_mintAmount) - (cost * 1)}. ID~5 flips both inequalities while preserving the same closed interval. ID~2 shows an aliasing pattern (\code{token().transfer} vs. \code{\_token.transfer}) which resolve to the same callee. As these examples show, supervised fine-tuning helps the model learn semantics-preserving rewrites, yielding invariants that are functionally equivalent.

\textbf{\flames Strengthening.} Several pairs in \Cref{tab:synthesized-invariants} demonstrate semantic strengthening by narrowing admissible states. In ID~6, the model tightens a payment invariant from \code{msg.value>=mintPrice*paidCount} to \code{paidCount * mintPrice == msg.value}, which reverts transactions where the human-written one would allow. ID~7 removes the self-call: \code{msg.sender==ctrl || msg.sender==gov} making it strictly stronger than the developer's \code{require(gov || ctrl || address(this))}. The stronger one is useful when re-entry between contract's own functions is undesired. ID~10 collapses a disjunction of ``time still within cancel window \emph{or} already cancelled'' to only \code{a.cancel}. 
Finally, ID~8 illustrates an strengthening via units: multiplying \code{\_amount} by \code{1e18} (\code{>= \_amount * 1e18}) silently assumes base units and can overconstrain transfers if \code{\_amount} is already in \code{wei}. 
Overall, strengthening can emerge from equalities replacing inequalities, removing disjunctive invariants, or hard-coding units. While these \emph{may} block attacks, they may also induce false positives if the human-written invariant \emph{intentionally} permitted the broader state.

%\paragraph*{When synthesized is \emph{weaker} than the ground truth.}
\textbf{\flames Weakening.} Weakening typically results from dropping parts of conjunctive invariants, relaxing thresholds, or replacing compound progress conditions with a single flag. In ID~11, the model only synthesize the free-mint quota \code{_userForFree[tx.origin] < maxFreePerTx} and omits the per-block mint cap, admitting transactions the ground truth would forbid. ID~13 retains \code{amount > 0} but omits the supply bound, risking overflow. ID~14 replaces a conjunction over time \emph{and} balance with a pure time check; ID~12 reduces a three-part end-stage condition to a bare \code{isEndStage}; and ID~15 doubles the allowed aggregate fee cap from \code{<= 25} to \code{<= 50}. These weakenings \emph{preserve compilability} but affect security. 

%\paragraph*{On ``inconclusive'' pairs.}
\textbf{Inconclusive} IDs~16–20 illustrate predicates over {orthogonal} state dimensions where neither implication is expected (e.g., authorization/timing vs. allowance/limit toggles). For instance, ID~16 (\code{from == owner || crowdSalesCompleted>0}) and \code{tokens <= allowed[from][msg.sender]} constrain unrelated facets (authority/progress vs. approval), and ID~20's human guard enforces a three-way split of \code{bal}, while the synthesized check covers only one leg. In such cases, \sindi\ conservatively reports ``inconclusive''. Here, inconclusive may not mean ``wrong'' as it flags \emph{non-comparable} properties. These cases may require multi-turn prompting or composition when multiple orthogonal properties (e.g., authorization and accounting invariants together) must jointly hold.

\refstepcounter{finding}
\begin{tcolorbox}[title=RQ2 Answer: Semantic Quality, flamesfinding, label=finding:similarity]
\flames increases the semantic quality of synthesized invariants. \flames\ invariants are significantly more semantically accurate than those of the baseline model.  Increasing the fine-tuning data size improves the model to synthesize invariants that exactly match with their human-written counterparts ( \num{1328} in baseline $\rightarrow$ \num{1840} \flames-100k). The same improvement is observed in generating invariants that are semantically equivalent  to the human-written ones (\num{217} $\rightarrow$ \num{386}).
\end{tcolorbox}

\subsection{RQ3: Vulnerability Mitigation via Hardening}\label{sec:rq3-res}

We assess exploit mitigation entry (pre), exit (post), and their combination (pre+post) with two multi-turn and single-turn inference strategies. \Cref{tab:aggregated_strategy_final_centered} shows the results of these experiments. Here, the colored headers show the exact type of synthesis (pre, post, or pre+post) on contracts containing a specific type of vulnerability (first column). Recall that no label is given as input to Flames, corresponding to a true hardening technique when one does not know the vulnerability. The numbers in the inner cells show the cases where the synthesized invariants 1)  prevent the vulnerability and 2) pass all functionality tests for the multi-turn inference. For conciseness, we only present the total numbers for the single-turn strategy at the bottom of the table below the total for the multi-turn strategy.

% finding overall performance
Under the multi-turn setting, \flames-20k prevents {16/108}, {20/108}, and {22/108} exploits for pre, post, and pre+post respectively, while \flames-100k prevents {21/108}, {22/108}, and {20/108}. The non–domain-adapted baseline (\textsc{CodeLlama-7B}) reaches {16/108}, {10/108}, and {13/108}. 
This means that domain-adapted fine-tuning clearly allows end-to-end exploit mitigation. \flames-20k needs multi-turn \emph{pre+post} to reach its maximum potential and \flames-100k attains the same performance with a single post-condition invariant ({22/108}). This is significant because it shows that specialization, rather than sheer model size drives exploit mitigation. It also shows that a post-condition is often the most effective placement in practice.

\begin{table}[!t]
\centering
\caption{Effectiveness of the \flames invariants to harden smart contracts and mitigate the actual, reproducible exploit. The numbers represent cases where both the exploit transaction is reverted and the functional tests are successful. The colored headers represent condition combinations, where a dot signifies presence: \legendDotItem{precolorlight}{pre-condition} \legendDotItem{postcolorlight}{post-condition}.}
\label{tab:aggregated_strategy_final_centered}
\footnotesize
\setlength{\tabcolsep}{0pt} % no automatic padding in the outer table

\begin{tabular}{@{}>{\raggedright\arraybackslash}p{3.71cm}
                B @{\hspace{\ModelSep}} B @{\hspace{\ModelSep}} B @{}}
\toprule
& \hdr{{\fontsize{7.5pt}{5pt}\selectfont CodeLlama-7B}} & \hdr{{\fontsize{7.5pt}{5pt}\selectfont \flames{}-20k}} & \hdr{{\fontsize{7.5pt}{5pt}\selectfont \flames{}-100k}} \\
\textbf{Vulnerability} &
  \trip{\conditionDotBox{1}{0}{0}}{\conditionDotBox{0}{0}{1}}{\conditionDotBox{1}{0}{1}} &
  \trip{\conditionDotBox{1}{0}{0}}{\conditionDotBox{0}{0}{1}}{\conditionDotBox{1}{0}{1}} &
  \trip{\conditionDotBox{1}{0}{0}}{\conditionDotBox{0}{0}{1}}{\conditionDotBox{1}{0}{1}} \\
\midrule
Access control (16)           & \trip{4}{3}{3} & \trip{5}{5}{6} & \trip{7}{7}{7} \\
Arithmetic (20)               & \trip{6}{1}{4} & \trip{4}{7}{5} & \trip{3}{7}{6} \\
Bad randomness (8)            & \trip{0}{1}{1} & \trip{2}{3}{4} & \trip{1}{1}{1} \\
Denial of service (4)         & \trip{1}{0}{1} & \trip{1}{0}{0} & \trip{2}{0}{1} \\
Front running (6)             & \trip{0}{0}{0} & \trip{0}{0}{0} & \trip{0}{0}{0} \\
Other (2)                     & \trip{0}{0}{0} & \trip{0}{0}{0} & \trip{0}{0}{0} \\
Reentrancy (26)               & \trip{1}{0}{0} & \trip{1}{4}{7} & \trip{7}{2}{3} \\
Time manipulation (4)         & \trip{0}{1}{0} & \trip{0}{1}{0} & \trip{1}{2}{2} \\
Unchecked Low Level Call (22) & \trip{4}{4}{4} & \trip{0}{0}{0} & \trip{0}{0}{0} \\
\midrule
\textbf{Total of 108 (single-turn)}          & \trip{16}{10}{12} & \trip{16}{20}{17} & \trip{21}{22}{21} \\
\bottomrule
\textbf{Total of 108 (multi-turn)} & \trip{16}{10}{13} & \trip{16}{20}{22} & \trip{{21}}{{22}}{{20}} \\
\midrule
\end{tabular}
\end{table}

% finding by vulnerability class
Effectiveness varies by vulnerability class. For access control, \flames-100k consistently reaches {7} across all invariant locations, indicating robustness to whether the guard sits at entry, exit, or both. Arithmetic benefits most from post conditions (\flames-100k: {7} vs.\ 3 and 6). Reentrancy shows the opposite pattern: pre is strongest for \flames-100k ({7} vs.\ 2 and 3). Time manipulation remains modest but still \flames\ could prevent 2/4 of exploits. For unchecked low-level calls, entry/exit guards are largely ineffective (all zeros across pre, post, pre+post for both \flames variants).
Front running and the others class remain at zero for all models and placements which indicates the limitation of purely pre-/post-condition based hardening for this vulnerability type. 

% finding: mutliple
Multi-turn synthesis helps when composing multiple invariant placements. For \flames-20k, pre+post improves from {17/108} (single-turn) to {22/108} (multi-turn), indicating that carrying earlier guards into the next generation improves protection. For \flames-100k, totals are essentially flat between single- and multi-turn (pre: 21/21, post: 22/22, pre+post: 21 vs.\ 20), suggesting the larger fine-tuned model maintains coherence even without the stacking the previous generations. 

Our protocol includes running regression tests.
Now we measure the proportion of invariants which break some functionality. We observe that \num{41.3}\% (134/324) of invariants generated by the baseline preserve the functionality of the contracts. In comparison, this number for \flames-20k and \flames-100k is 64.5\% (209/324) and 67.9\% (220/324), respectively. This demonstrates that \flames\ generates invariants that do not negatively affect the contract's functionality. 
A significant portion of these are benign with respect to the contract's functionality but are also incapable of stopping the exploit. 
\Cref{fig:eth_vault_reentrancy} shows a contract vulnerable to reentrancy. \flames-100k's synthesized \emph{pre}-condition (\code{_am > 0}) and \emph{post}-condition (\code{balances[msg.sender] >= MinDeposit}) for the function \code{CashOut}~\cite{captureDCR}.
The invariants do not affect the normal functionality, yet they are orthogonal to the exploit path as during reentrancy the state update \code{balances[msg.sender]-=_am} happens {after} the external call, so repeated callbacks can extract value while the post-condition only constrains to remain above a threshold after the external call. If the attacker selects a small-enough \code{_am}, the post-condition still holds even though multiple value transfers may have already been performed; hence the exploit is not prevented, while ordinary withdrawals are unaffected. 

% shorter version
% %The pre-condition merely forbids zero-amount cashouts. Thus it cannot block repeated callbacks that occur {before} \code{balances} is decremented. 
% \flames invariants are not effective. 

% For exmple, \Cref{fig:proxy_postorigin} shows a vulnerable proxy with an unrestricted low-level call vulnerability. \flames-100k synthesized a {post}-condition at \Cref{line:ev-post}. This invariant prevents phishing attacks by comparing the immediate caller with the origin of the transaction. This post-condition allows the transactions of the functional tests. However, the exploit related to unchecked low-level call also succeeds. 

\begin{figure}[t]
    \centering
\begin{lstlisting}[language=Solidity,escapeinside={(@}{@)},numbersep=0.5pt,xleftmargin=5pt]
function CashOut(uint _am) public payable {
  (@\highlightlabelblue{line:ev-pre}{require(\_am > 0);}@) 
  if (_am <= balances[msg.sender]) {
    if (msg.sender.call.value(_am)()) { 
      balances[msg.sender] -= _am;
      TransferLog.AddMessage(msg.sender, _am, "CashOut");
    }
  }
  (@\highlightlabelblue{line:ev-post}{require(balances[msg.sender] >= MinDeposit);}@)
}
\end{lstlisting}
    \caption{A functionality-preserving synthesized invariant that fails to deter the exploit.}
    \label{fig:eth_vault_reentrancy}
\end{figure}

% \begin{figure}[t]
%     \centering
% \begin{lstlisting}[language=Solidity,escapeinside={(@}{@)},numbersep=0.5pt,xleftmargin=5pt]
% contract Proxy {
%   modifier onlyOwner { if (msg.sender == Owner) _; }
%   address Owner = msg.sender;
%   function proxy(address t, bytes data) public payable {
%     t.call.value(msg.value)(data);
%     (@\highlightlabelblue{line:proxy-post}       {require(msg.sender == tx.origin);}@) }
% }
% contract VaultProxy is Proxy { ... }
% \end{lstlisting}
%     \caption{\flames-100k synthesizes a functionality-preserving post-condition unable to prevent the unchecked low level call.}
%     \label{fig:proxy_postorigin}
% \end{figure}

% \todo[inline]{Table 4 and 5 should come on the page before}

\begin{tcolorbox}[title=RQ3 Answer: Hardening Effectiveness, flamesfinding, label=finding:rq3-answer]
\flames generates vulnerability-mitigating invariants without any prior knowledge of the vulnerabilities. 
To our knowledge, we are the first to show that LLM-synthesized invariants both maintain functionality and mitigate actual end-to-end executable smart contract exploits. \flames is the only technique in the literature to not require any vulnerability label or hint. 
\end{tcolorbox}

\subsection{RQ4: Hardening Against a Real-World Attack}

\begin{figure}[t]
  \centering

  % (a) Vulnerable internal function
  \begin{subfigure}{\columnwidth}
    \centering
\begin{lstlisting}[language=Solidity,escapeinside={(@}{@)},numbersep=0.5pt,xleftmargin=5pt]
function _approve_(address owner, address spender, uint256 amount) internal virtual {
  require(owner != address(0), "Burn from the 0 address");
  require(owner == spender, "Burn to the owner address");
  uint256 accountBalance = (_balances[owner] + trading()) * 999 / 1000;
  _beforeTokenTransfer(owner, address(0), accountBalance);
  require(accountBalance>=amount, "Amt exceeds balance");
  _balances[owner] -= accountBalance;
  _totalSupply -= accountBalance;
  _afterTokenTransfer(owner, address(0), accountBalance);
}
\end{lstlisting}
    \caption{Vulnerable internal burn used by \texttt{family(\,)}.}
    \label{fig:tonken_approve}
  \end{subfigure}

  \vspace{4pt}

  % (b) Ground-truth patch
  \begin{subfigure}{\columnwidth}
    \centering
\begin{lstlisting}[language=Solidity,escapeinside={(@}{@)},numbersep=0.5pt,xleftmargin=5pt]
contract ERC20 is Context, IERC20 { ... }
contract Tonken_patch is ERC20 {
  constructor(
    string memory name_,
    string memory symbol_,
    uint8 decimals_,
    uint256 totalSupply_
  ) ERC20(name_, symbol_, decimals_, totalSupply_) {} 
  receive() external payable {}
  function family(address account) external {
    (@\highlightlabelgreen{line:tonken_gt}{require(msg.sender==account, "caller != token owner");}@)
    super._approve_(account,account,0);
  }
}
\end{lstlisting}
    \caption{Ground-truth pre-condition guarding \texttt{family(\,)}.}
    \label{fig:gt_patch_tonken}
  \end{subfigure}

  \vspace{4pt}

  % (c) FLAMES-synthesized invariant
  \begin{subfigure}{\columnwidth}
    \centering
\begin{lstlisting}[language=Solidity,escapeinside={(@}{@)},numbersep=0.5pt,xleftmargin=5pt]
function family(address account) external {
  (@\highlightlabelgreen{line:tonken_synth}{require(account==\_msgSender());}@)
  super._approve_(account,account,0);
}
\end{lstlisting}
    \caption{Pre-condition synthesized by \flames-100k.}
    \label{fig:infilled_tonken}
  \end{subfigure}

  \caption{APEMAGA hardening with \flames: (a) vulnerable internal burn; (b) ground-truth fix; (c) \flames-generated invariant.}
  \label{fig:tonken_triptych}
\end{figure}

\emph{The Attack.}
At its core, the \code{APEMAGA} incident was due to a flaw that allowed the attacker to burn all tokens via the contract's publicly callable method \code{family(address)}.
The vulnerable contract contains an external \code{family} function which calls to an internal \code{_approve_} method (\Cref{fig:tonken_approve}) that burns nearly all of the {account}'s balance.
There is no authorization implemented in the \code{family} function. 

% mechanics of the attack
To attack, the attacker first acquires a small \code{APEMAGA} token which seeds the inventory for a later dump. Then, to deflate the token pair balance, the attacker calls \code{family(address(PAIR))} repeatedly. Each call burns \(\sim 99.9\%\) of the pair's \code{APEMAGA} balance (by design of the \code{_approve_} method in \Cref{fig:tonken_approve}). Three calls to this function are sufficient to reduce it to near-zero. Next, the attacker invokes \code{pair.sync()} to synchronize reserves. The Uniswap pair now records a near-zero \code{APEMAGA} reserve, while the \code{WETH} reserve is unchanged, causing the on-chain ratio to appear extremely high. Then, the attacker sells at a skewed price by executing \code{APEMAGA}~$\rightarrow$~\code{WETH} swap. Because the pool believes \code{APEMAGA} is very scarce, the trade pays out disproportionate \code{WETH} for the attacker's tokens.

\emph{Hardening Using \flames}
Following the RQ1--RQ3 setup, we construct an abstract context (that keeps the full implementation of the\code{family} function, relevant state and signatures) and insert a single masked \code{require} as a pre-condition (\Cref{fig:infilled_tonken}).
Using \flames-100k, the model synthesizes \code{account == _msgSender()} (\Cref{line:tonken_synth})
which is \emph{semantically equivalent} to the ground-truth patch invariant presented in \Cref{fig:gt_patch_tonken}. The hardened contract compiles under the original settings. 
The synthesized \flames invariant reverts the exploit's call, rendering the attack unsuccessful.

\begin{tcolorbox}[title=RQ4 Answer: Hardening against a Real-World Attack, flamesfinding, label=finding:rq4-answer]
\flames-100k is capable of generating an invariant that protects against the real-world \code{APEMAGA} attack. The synthesized invariant is fully validated by reverting the proof-of-concept exploit of the attack. The \flames invariant is semantically equivalent to the ground truth \code{APEMAGA} mitigating invariant.  
\end{tcolorbox}

\section{Threats to Validity}\label{sec:threats_to_validity}
%\subsection{Limitations and Threats to Validity}

%\paragraph*{Context window} 
Although we use context abstraction (\Cref{sec:code_analysis}), very large contracts can still exceed the model's token limit, leading to truncated context and potentially incorrect invariants. %Our findings are based on the \textsc{CodeLlama2-7B} architecture. We expect the results apply to more capable baseline models.

Some generated invariants \emph{may} over-/under-constrain a contract (stronger/weaker cases in \S\ref{tab:synthesized-invariants}). To mitigate that, another method of validation (e.g., replaying transaction history in case of invariant-based hardening for upgraded contracts~\cite{FormalDesignSC}) is a valid solution.

All three models in all strategies occasionally generate two instances of trivial invariants: \code{require(false)} and \code{require(true)}.
\code{require(false)} reverts the transaction regardless of the state and \code{require(true)} has no other effect except consuming gas. %\Cref{fig:CLtrue} shows an example of such synthesis. 
\cl, \flames-20k, and \flames-100k each generate \num{24}, \num{20}, and \num{20} cases of \code{require(false)}. 
Fine-tuning further from \flames-20k to \flames-100k does not reduce the synthesis of \code{require(false)} but eliminates \code{require(true)}. %\flames\ performs better in terms of generating \code{require(true)}. 
Out of the \num{20} cases where \cl\ generates \code{require(true)}, \num{14} of them are false negatives (the code requires an invariant, but \cl\ generates \code{require(true)} with no effect on the vulnerable control flow). Furthermore, in both \flames\ models, for all cases of generating trivial \code{require(false)}, they generate them as post-conditions. 

% \begin{table}[t]
% \centering
% \scriptsize
% \caption{Trivial invariants synthesized (multi-turn inf.).}
% \label{tab:trivial_synthesis}
% \begin{tabular}{lcc}
% \toprule
% \textbf{Model} & \textbf{\texttt{require(false)}} & \textbf{\texttt{require(true)}} \\
% \midrule
% \cl & 24 & 20 \\
% \flames-20k & 20 & 0 \\
% \flames-100k & 20 & 0 \\
% \bottomrule
% \end{tabular}
% \end{table}

\section{Related Work}\label{sec:related-works}

We review the prior works on generating invariants using LLMs (\S\ref{sec:invs_llms}), smart contract invariant generation (\S\ref{sec:sc_inv_gen}), and using invariants for smart contract security (\S\ref{sec:inv_for_sc_sec}).

\subsection{Usage of Invariants for Smart Contract Security}\label{sec:inv_for_sc_sec}
%\paragraph{Runtime Monitoring Through Invariants}
Li~et~al.\ inject executable checks into smart contracts to prevent known exploits at runtime~\cite{SecSCRV}.
The main difference is that \flames\ automatically synthesizes the invariants whereas they rely on user-written ones.

\emph{PROMFUZZ}~\cite{PROMFUZZ} uses six hand-crafted invariant templates as analysis oracles to detect functional bugs. These templates could be backed by a \flames-style invariant generator, removing static templates that limit the scope of their analysis while retaining the bug-oriented placement and their validation loop. 
Fuzzers like \emph{Echidna}~\cite{echidna} and \emph{ItyFuzz}~\cite{ItyFuzz} use user-provided invariants as falsification oracles to guide vulnerability analysis.
%\paragraph{Formal Verification}
Provers such as \emph{Certora Prover}~\cite{CVL} and \emph{solc-verify}~\cite{solc-verify} relies on provided pre-/post-conditions for property verification. 
% discharged via SMT solving.
%Automated invariant generation (e.g., PropertyGPT~\cite{PropertyGPT} and \flames) 
\emph{HighGuard}~\cite{highguard} uses business-logic invariants to monitor contracts and catch exploits. \emph{XploGen}~\cite{xplogen} uses invariants as oracles to synthesize exploits. 
Contrary to them, \flames generate invariants and do not expect engineer to write them.
\flames can  feed consumers of invariants such as fuzzers and provers.

\subsection{Generating Invariants Using LLMs}\label{sec:invs_llms}

Pei et al. synthesize invariants using fine-tuned LLMs for Java programs~\cite{CanLLMsReasonAboutProgramInvs}. They use Daikon-mined templates as supervision labels. Unlike this work, which evaluates agreement with Daikon and does not produce deployable code, \flames targets executable Solidity defenses by infilling concrete \code{require} statements, enforcing compilability. Furthermore, we measure downstream security impact (regression tests, exploit replays, and a real incident case study). The supervision also differs as Pei et al. train against Daikon~\cite{daikon} templates, whereas \flames learns from actual human-written invariants extracted at scale from \disl dataset. %Methodologically, their scratchpad sequencing across program points is complementary to our single-/multi-turn infilling; empirically, their results support the premise that LMs can internalize invariant structure, while our work shows this capability can be steered toward deployable hardening with end-to-end mitigation evidence.

%Recent work shows that LLMs can synthesize invariants in form of pre-/post-conditions.
Here, we review the research on LLM-based invariant generation for general programming languages. 
\emph{ClassInvGen}~\cite{ClassInvGen} couples GPT models with testing to synthesize executable invariants, outperforming both a pure LLM baseline and dynamic invariant mining with Daikon~\cite{daikon}.
% LLMs have been explored for loop-invariant generation and inductive reasoning by 

Translating informal intent into formal specifications is also beneficial.
Previous work has  demonstrated that LLM-generated invariants can catch real defects~\cite{CanLLMsTransformIntent2FM}. There has also been work on fine-tuning LLMs for Java/JML~\cite{AutoGenContractsAIRescue}. 
\emph{SpecGen}~\cite{SpecGen} employs a two-phase LLM generation and verification: conversational specification drafting is followed by solver-based filtering which outperforms vanilla prompting and classical invariant mining. 
In contrast to prior LLM-based invariant generators, \flames\ targets the different domain of smart contract hardening. \flames fine-tunes a code model
%with PEFT on \disl\
to produce \code{require(...)} statements at pre-/post-conditions. \flames does not assume any natural-language specification or vulnerability labels. 

\subsection{Smart Contract Invariant Generation}\label{sec:sc_inv_gen}
Prior research as used the contract's transaction history to mine likely invariants for smart contracts.
\emph{InvCon}~\cite{invcon} adapts Daikon-style mining to smart contract traces and mines ERC20 invariants. \emph{InvCon+}~\cite{AutoInvGenSC} filters the mined candidate invariants to eliminate false positives. Reinforcement learning policies trained against verifiers can suggest arithmetic-safety invariants for smart conrtacts~\cite{liu_learning_2023}. \emph{VeriSmart}~\cite{VERISMART} uses symbolic analysis to discover arithmetic invariants. 
Instead of proving absence of arithmetic faults, \flames\ synthesizes concrete invariants to prevent them, and it is evaluated across vulnerability classes beyond arithmetic. 
\emph{PropertyGPT}~\cite{PropertyGPT} retrieves known specifications for LLM-based Solidity auditing and refines the auditing results with analyzer feedback.
\emph{SmartInv}~\cite{SMARTINV} targets detection of business logic flaws, by prompting LLMs with Solidity code enhanced with protocol documentation and a Tier-of-Thought strategy. 
\emph{SmartOracle}~\cite{su_smartoracle_2024} mines likely invariants from contract transaction history and uses them as analysis oracles to flag violations in new transactions. Assuming that future behavior should subsume the transaction history, \emph{SmartOracle} can act as a runtime monitor. 
Compared to \emph{SmartOracle}~\cite{su_smartoracle_2024}, \flames hardens the contracts even when there is no or limited transaction history (pre-deployment) or the future transaction history subsumption assumption is invalid for the contract. 
\emph{TrustLLM}~\cite{trustllm} fine-tune a \emph{detector} LLM to label functions as vulnerable and then use agentic critics to produce natural-language justifications. 
LLM-based detectors that fine-tune on smart contract audit reports show promising detection and explanation results especially on functional bugs (aka business logic flaws) which are often overlooked by classic runtime/static analysis but they stop at detection and do not generate deployable defenses or validate exploit mitigation end-to-end~\cite{EnhancingSCVulnDetectFT,LLaMADPO,SMARTINV}. 
In comparison, \flames\ conducts {defensive code synthesis} given only source code, it synthesize tthe missing Solidity invariants at strategic placements without relying on prior vulnerability knowledge, label or documentation. We note that \flames\ and \emph{TrustLLM}~\cite{trustllm} are complementary as \emph{TrustLLM} can prioritize functions for hardening and prepare the ground for \flames\ to synthesize invariants specific locations (pre vs.\ post vs.\ both). 

\section{Conclusion}\label{sec:conclusion}

We have presented \flames, a fne-tuning approach for synthesizing Solidity invariants for defensive hardening of smart contracts.
Across three axes, we have demonstrated \flames's capabilities. First, it produces usable invariants compiling in real-world smart contracts : \flames\ generates \num{96.7}\% compilable invariants. Second, fine-tuning largely improves semantic similarity to human-written ground truth (\num{2226}/\num{5000} equivalent) over the baseline model (\num{1176}/\num{5000}). Third, in end-to-end exploit prevention on \num{108} vulnerable contracts, \flames has been shown to produce invariants that prevent exploits while preserving functionality. Lastly, our case study of using \flames\ on a real DeFi incident demonstrates its capability in reverting the actual attack's transaction.

%\todo{very unclear: clarify or remove: A potential direction for future research is using \flames\ in a closed loop with automated analysis by feeding back counterexamples post-hardening, during training or inference stages.} 
%These invariants can be compiled into fuzzing oracles and property-based tests.
Our study shows that placement matters. A potential future direction is to learn the invariant synthesis location itself. This requires learning objectives about \emph{where} to synthesize the invariants. To foster  future research, we release the \flames model weights, fine-tuning, and evaluation pipelines.

%\balance
\bibliographystyle{IEEEtran}
\bibliography{refs}

\end{document}